# Effect of nuclear deformation on direct capture reactions

FAN Guang-Wei(樊广伟)[1,*]  CAI Xiao-Lu(蔡晓鹭)[2]  M. Fukuda[3]
HAN Ti-Fei(韩体飞)[1]  LI Xue-Chao(李学超)[1]  REN Zhong-Zhou(任中洲)[4]
XU Wang(徐望)[2]

[1] (School Of Chemical Engineering, Anhui University of Science and Technology, Huainan, 232001, China)
[2] (Shanghai Institute of Applied Physics, Chinese Academy of Sciences, Shanghai 201800, China)
[3] (Department of Physics, Osaka University, Osaka 560-0043, Japan)
[4] (Department of Physics, Nanjing University, Nanjing 210008, China)

**Abstract**：The direct radiative capture process is well described by the spherical potential model. In order for the model to explain direct captures more accurately, the effect of the nuclear deformation has been added and analyzed in this work, since most nucleuses are not spherical. The results imply that the nuclear deformation largely affects the direct capture and should be taken into account during discussing direct capture reactions.

## 1. Introduction

Low-energy nuclear direct capture reactions play a crucial role in big bang nucleosynthesis, main path stellar evolution, element synthesis at supernova sites, X-ray bursts etc., since cross sections of the nuclear reactions are often necessary for investigating the astrophysical entities [1-3]. However, some reactions occur at energies that are almost not directly accessible in terrestrial laboratories. Furthermore, some reactions are practically impossible to be directly measured, such as the $^{8}$Li(n, γ)$^{9}$Li capture reaction, because there is no $^{8}$Li or neutron target exist [4]. These necessitate theoretical extrapolation from higher to lower energies or pure theoretical calculations with no firm experimental basis.

The direct capture process represents a transition for the projectile from an initial continuum state to a final bound state via interaction with the electromagnetic field. The reaction selects those projectiles from the appropriate partial waves with orbital angular which can jump into final orbits by emitting γ-ray of multipolarity *L*. In order to calculate the direct capture cross sections, one needs to solve the many-body problems for the bound and continuum states of relevance for the capture process. There are several levels of difficulty in attacking this problem. The simplest solution, among theories which have been developed to overcome the difficulties, such as microscopic cluster model [5], R-matrix [6] etc., is based on a potential model to solve the initial and final state. The model treats the direct capture reaction as a core plus a valence part interacting via a potential, for example Woods-Saxon (WS) potential. Thus, cross sections are sensitive to the potential, for wave functions (WFs) are sensitive to the shape of the potential [7-9].

Because the nuclear deformation largely affects its corresponding potential, and most nucleuses are not spherical, for instance, famous nucleuses $^{8}$Li and $^{8}$B ($^{7}$Li(n, γ)$^{8}$Li and $^{7}$Be(p, γ)$^{8}$B) [10]. We will add the nuclear deformation to the potential model and analyze effects of the nuclear deformation by calculating the cross sections of the $^{7}$Li(n, γ)$^{8}$Li reaction in this article. In the following section, brief framework of the potential model and the nuclear deformation considered are presented. Deformed effects analyzed by calculating cross sections of the $^7$Li(n, γ)$^8$Li reaction is shown in Section 3, followed by a summary in Section 4.

## 2. Framework of the potential model

Since the potential model is a standard model, we will briefly outline the formalism of this theory in this paper. The details are referred to Refs. [11,12]. Within the potential model, the capture cross section of the capture reaction $a(x, γ)b$ is given by

$$\sigma_{a \to b}^{(EL)} = \frac{8\pi(L+1)}{L\left[(2L+1)!!\right]^2} \frac{k_\gamma^{2L+1}}{\hbar v} \frac{1}{2s+1} \frac{1}{2I+1} \times e_{EL}^2 \sum \left|Q_{a \to b}^{(EL)}\right|^2, \tag{1}$$

where $k_\gamma = \varepsilon_\gamma / \eta c$ is the wave number for a transition emitting a γ-ray with energy $\varepsilon_\gamma$, $e = -Z/A$ is the effective charge for neutrons. The matrix elements for electromagnetic transitions of electric multipolarity $L$ is expressed as

$$Q_{a \to b}^{(EL)} = \left\langle \Psi_a \left\| T^{EL} \right\| \Psi_b \right\rangle, \tag{2}$$

where $T^{EL} = r^L Y_{LM}$ stands for the electric multipole operator. The matrix elements can be explained as a product of three factors

$$Q_{a \to b}^{(EL)} = \tau_{a,b}^{EL} B_b A_{a,b}, \tag{3}$$

$$\tau_{a,b}^{EL} = \int \psi_{scat} r^L \psi_{bound} dr, \tag{4}$$

where the equation (4) denotes the overlap integral between the radial components of the continuum or incoming particle WF $\psi_{scat}$ scattered by the $a$-x potential and bound state WFs $\psi_{bound}$ of b. The quantity $B_b$ represents the fractional parentage coefficient for a single-particle configuration of the bound state (spectroscopic factor) and $A_{a,b}$ denotes an angular momentum coupling coefficient.

For charged particles the astrophysical $S$-factor for the direct capture from a continuum state to the bound state is defined as

$$S^{(c)}(E) = E\sigma_{a \to b}^{(EL)}(E) \exp\left[2\pi\eta(E)\right], \text{ with } \eta(E) = Z_a Z_b e^2 / \hbar v, \tag{5}$$

where $v$ is the relative velocity between $a$ and b.

WFs $\psi_{scat}$ and $\psi_{bound}$ in the potential model are obtained by solving the scattering and bound-state systems, respectively, for a given potential. Thus, the essential ingredients of the model are potentials used to generate the WFs, especially, the potential to generate $\psi_{bound}$, since the continuum state can be determined precisely from the elastic scattering experiment. The potentials used in the potential model are spherical, such as WS potential. This may cause potential uncertainties in calculating cross sections, for most nucleuses are not spherical. Therefore, we modify the potential model with deformed WS potential by instead of spherical one. The spherical WS potential is pressed as

$$V = \left(-V_0 + V_1(l \cdot s)\frac{r_{l \cdot s}^2}{r}\frac{d}{dr}\right)f(r), \tag{6}$$

Where $f(r) = [1+\exp((r-R_c)/a_0)]$, $a_0$, the diffuseness parameter, $R_c$, the radius of the potential, $r_{l*s}$,

the radius for spin-orbit potential [9,12]. The depth of $V_0$ is adjusted to reproduce the experimental binding energy of the valence part of nucleus. The deformed WS potential is obtained by modifying f(r) with

$$f(r) = 1/[1 + q \exp((r - R_c) / a_0)] \qquad (7)$$

where $q(= 0.6)$ is the real parameter which determines the shape (deformation) of the potential [13].

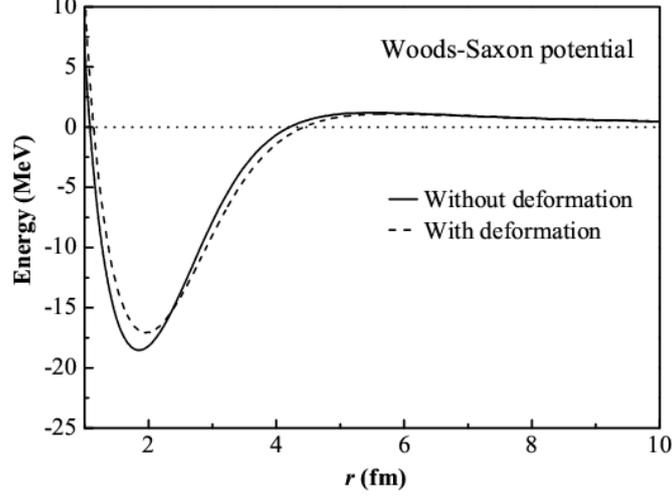

Fig. 1, The difference of potentials given by spherical and deformed WS.

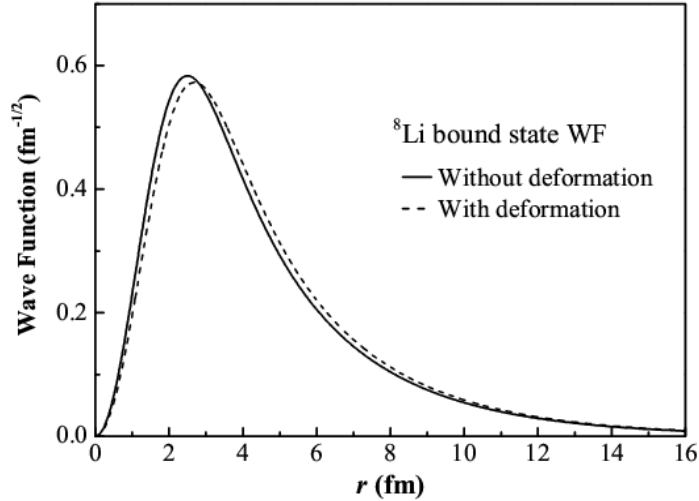

Fig. 2, The difference of the valence bound-state WF of $^8$Li between with and without deformation correction.

## 3. Calculations of $^7$Li(n, γ)$^8$Li reaction

$^7$Li(n, γ)$^8$Li reaction is a key reaction in inhomogeneous big bang nucleosynthesis to jump the $A$ = 8 gap. It is representative in theoretical studies. For $^7$Li(n, γ)$^8$Li, the direct radiative capture of a s- or d- wave neutron by $^7$Li, leaving the $^8$Li compound nucleus in either the g.s. ($J^\pi = 2^+$) or the first excited state ($J^\pi = 1^+$) proceeds by an $E1$ transition. To discuss the reaction, the spherical and deformed WS potential formed with $r_0 = 1.25$ fm, $a_0 = 0.65$ fm are adopted. The depth of the potentials $V_0$ ($V_0$(g.s.) and $V_0$(1st)) are adjusted to reproduce the corresponding binding energies, $E_{gs} = 2.033$ MeV and $E_{1st} = 1.052$ MeV. The potential (spherical and deformed) used to describe

the scattering of the neutron by $^7$Li also has geometric parameters $r_0 = 1.25$ fm and $a_0 = 0.65$ fm. The well depth has been adjusted in order to reproduce the experimental scattering length $a_+ = -3.63 \pm 0.05$ fm and $a_- = 0.87 \pm 0.05$ fm for the two components of the channel spin s at thermal energies. The results of this analysis, assuming only an s-wave capture, are compared in Fig. 3.

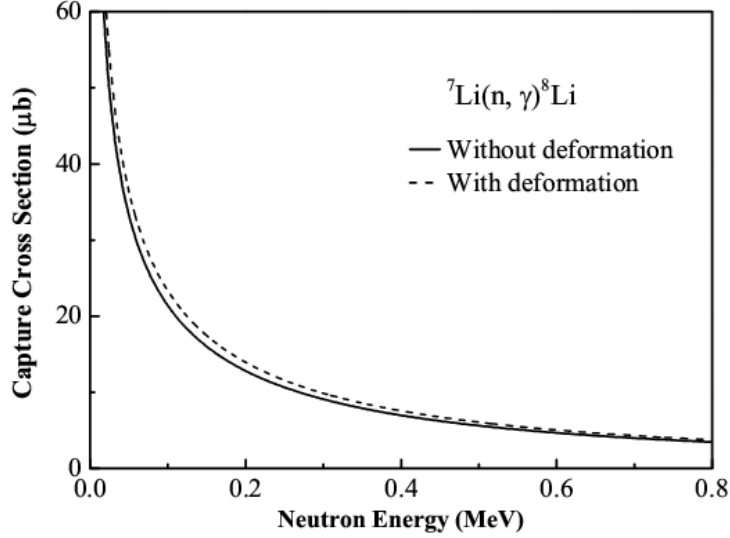

Fig. 3, The cross sections for the $^7$Li(n, γ)$^8$Li as a function of neutron energy.

From figures above, one can see that nuclear deformation largely affects the shape of the potential and thereby WF of the valence part and even the cross sections. It supports the conclusion of Y. Nagai *et al.* [9] that the cross sections are sensitive to the interaction potential. From the analysis, the difference of the cross sections by two kinds of the potential up to 7.8%, this will greatly change the components of the nucleosynthesis. As a result, the deformation obviously affects the amount of heavier elements produced in the inhomogeneous big bang theory. Thus, the nuclear deformation should be taken into account for potential model in extrapolations. In this work, we wanted to give the cross sections for $^7$Li(n, γ)$^8$Li reaction within the deformation effect, unfortunately, it is suspended because there is no precise range $R_c$ and diffuseness parameter $a_0$ for the deformed WS potential, since different choices for $R_c$ and $a_0$ can be made to reproduce the binding energy, which can cause an irreducible uncertainty in the calculation. In nest article, we will discuss how to remove the irreducible uncertainty through deformed WS potential.

## 4. Summary

The authors have analyzed the effect of the nuclear deformation in the direct capture reactions through the key reaction of $^7$Li(n, γ)$^8$Li. The analyzed results show us that the nuclear deformation largely affects the cross sections and should be taken into account during discussing the direct capture reactions.

## Acknowledgements

The authors download the code of potential model from Computer Physics Community (http://cpc.cs.qub.ac.uk/summaries/ADSH), the analysis is based on the code. We would like to acknowledge the financial support provided by Anhui University of Science and Technology (11130).

**References**

[1] L. H. Kawano et al., Astrophys. J 372 (1991) 1.

[2] S. Burles et al., Phys. Rev. Lett. 82 (1999) 4176.

[3] C. Rolfs et al., Cauldrons in the Cosmos (University of Chicago Press, Chicago, 1988).

[4] V. Guimarães et al., Phys. Rev. C 75 (2007) 054602.

[5] P. Descouvemont and D. Baye, Nucl. Phys. A 487 (1988) 420.

[6] R. M. Kremer et al., Phys. Rev. Lett. 60 (1988) 1475, and references therein.

[7] T. Tombrello, Nucl. Phys. 71 (1965) 459.

[8] B. Davids and S. Typel, Phys. Rev. C 68 (2003) 045802.

[9] Y. Nagai et al., Phys. Rev. C 71 (2005) 055803.

[10] Gautam Rupak and Renato Higa, Phys. Rev. Lett. 106 (2011) 222501.

[11] C. Rolfs, Nucl. Phys. A 217 (1973) 29.

[12] C. A. Bertulani, Computer Physics Commun. 156 (2003) 123.

[13] Falaye. B. J, Hamzavi M and Ikhdair S M, http://arxiv.org/pdf/1207.1218.pdf